\documentclass[11pt,twoside]{article}
\usepackage[pdftex]{graphicx}
\usepackage{amsmath}
\usepackage{amssymb}
\usepackage{cite}

\usepackage[top=2cm, left=2cm, right=2cm, bottom=2cm]{geometry}

\usepackage[space-before-unit=true, use-xspace=true]{siunitx}
\usepackage{subfigure}
\usepackage{authblk}

\title{Correlations of multiplexed quantum ghost images and improvement of the quality of restored image\footnote{Published as Balakin,\,D.A., Belinsky,\,A.\,V., Chirkin,\,A.\,S. Correlations of multiplexed quantum ghost images and improvement of the quality of restored image // Journal of Russian Laser Research. 2017. Vol.~38, N~2. P.~164--172.}}
\date{}
\author[1]{Dmitriy Balakin}
\author[1]{Alexander Belinsky}
\author[1,2,*]{Anatoly S. Chirkin}

\affil[1]{M.\,V.\,Lomonosov Moscow State University, Faculty of Physics\newline
Leninskie Gory, 1, bld 2, Moscow 119991, Russia}
\affil[2]{M.\,V.\,Lomonosov Moscow State University, The International Laser Center\newline
Leninskie Gory, 1, bld 62, Moscow 119991, Russia}
\affil[*]{Corresponding author e-mail:~~~aschirkin~@~rambler.ru}

\begin{document}
\newcommand{\pst}{\hspace*{1.5em}}

\newcommand{\vq}{\mathbf{q}}
\newcommand{\vrho}{\boldsymbol{\rho}}
\newcommand{\vr}{\mathbf{r}}
\newcommand{\vd}{\mathbf{d}}
\newcommand{\vR}{\mathbf{R}}
\newcommand{\bydef}{\overset{\text{def}}{=}}

\maketitle

\begin{abstract}\noindent
The currently used ghost image schemes traditionally involve two-mode entangled light states or incoherent radiation.
Here, application of four-mode entangled light states is considered.
It is shown that multiplexed ghost  images (MGI) formed by four-mode entangled quantum  light states have mutual spatial correlations determined by the 8th order field correlation functions.
A special algorithm to calculate high-order correlations of Bose operators was developed.
We also demonstrate that the accounting of MGI correlations  allows us to improve the quality of the restored image of an object when processing MGI by measurement reduction method.
Computer modelling of recovery of the image from MGI was carried out.
It is established that in the considered example the signal-to-noise ratio of the reduced ghost image is $4.6$ times higher than the best signal-to-noise ratio for the ghost images themselves.
\end{abstract}

\medskip


\section{Introduction}
\pst
When observing a ghost image (GI), information on an object is extracted by measuring the spatial correlation between photons propagating through an object or reflected by it, and photons of reference arm which have not interacted with the object.
Object photons are measured by a single-pixel or bucket photodetector that has no spatial resolution, while photons of reference arm
are registered by a scanning single-pixel photodetector or CCD matrix. As a result, the spatial correlation function that contains information on the object \cite{Klyshko} is measured.
By now a number of schemes of spatial images have been suggested and carried out (see reviews \cite{Shapiro,Erkman,Gatti} and works \cite{Lopaeva,Chirkin,Zhang, Chirkin2,Luo }). The GI technique was extended to X-ray range in experiments \cite{Yu, Pelliccia}, and its application in \si{\tera\hertz} range is discussed in \cite{Youngner}.
Recently it was shown \cite{Ryczkowski} that this technique can be applied to reconstruct temporal ``images'' (information) of rapidly varying signal in telecommunication systems. GI attracts interest in connection with a possibility of simplification of  illuminating  an object, obtaining the image of an object is possible using even a small number of photons. The important characteristics of restored image are its contrast and its signal-to-noise ratio. 

The aim of this work is to show that  mutual correlation of multiplexed ghost images (MGI) can be used to improve quality of the restored image  by means of measurement reduction method.  Restoration and analysis of MGI are illustrated by an example, in which MGI are formed by four-mode entangled quantum light states. It is shown that in this case the signal-to-noise ratio in the restored image can be improved manyfold.

\section{Multiplexing ghost imaging}
\begin{figure}
\includegraphics[width=\linewidth]{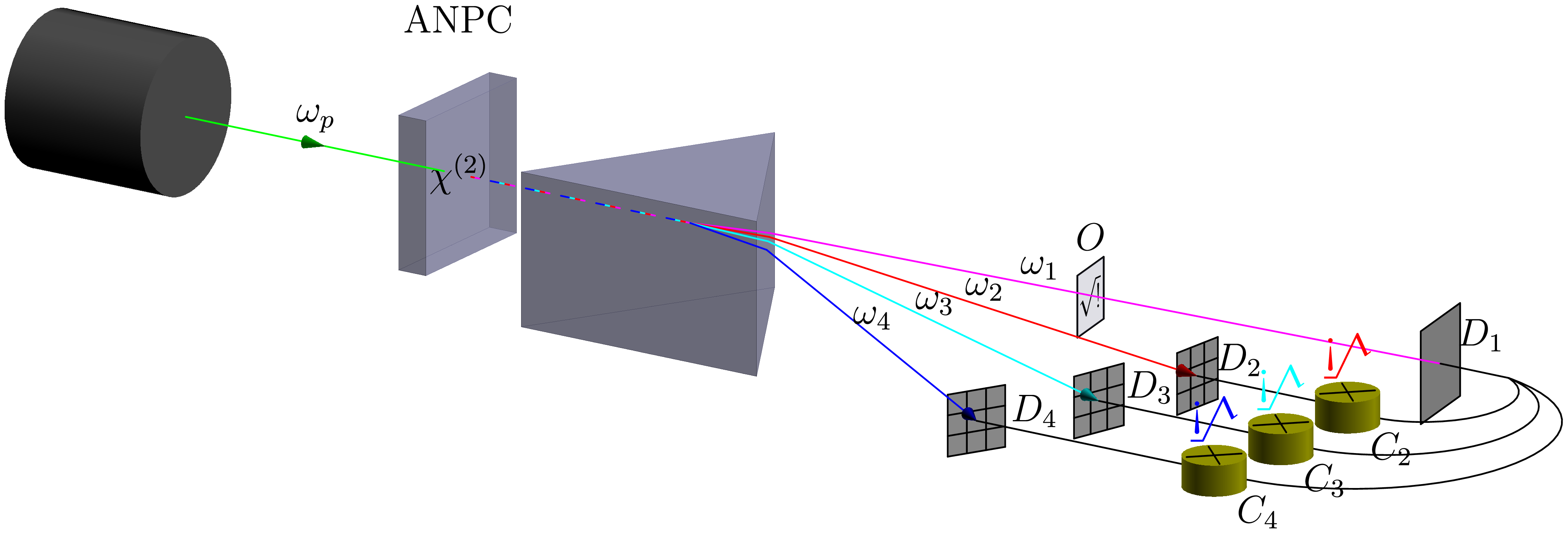}
\caption{The schematics of set-up of forming multiplicated ghost images: ANPC is an aperiodically polarized nonlinear crystal;
$\omega_p$ is pump frequency;
$\omega_{1}, \ldots, \omega_{4}$ are frequencies of entangled beams that are formed in ANPC;
$O$ is the object;
$D_{n}$ are the detectors in object ($n = 1$) and reference ($n = 2, \dots, 4$) arms;
$C_{n}$ are the correlators}
\label{fig:ghost-imaging}
\end{figure}

\pst
The schematics of set-up for obtaining MGI are shown in Fig.~\ref{fig:ghost-imaging}. Pump radiation, that is, intensive monochromatic radiation with frequency $\omega_p$,  falls on an aperiodical nonlinear photon crystal (ANPC). In the crystal,   pump photons are split into two photons with  frequencies related as
\begin{equation*}
\omega_p = \omega_1 + \omega_2.
\end{equation*}

Four-frequency field is obtained as a result of  subsequent conversion of a part of photons with frequencies 
$\omega_{1}$ and  $\omega_{2}$ into photons with frequencies
 $\omega_{3}$ and  $\omega_{4}$ in frequency mixing process:
\begin{eqnarray*}
\omega_{p} + \omega_{1} &= \omega_{3},\\
\omega_{p} + \omega_{2} &=  \omega_{4}.%
\end{eqnarray*}

Effective power exchange between the interacting light waves is carried out if quasi-phase matching condition is satisfied, that is,
phase mismatches $\Delta  k$ between
the interacting waves are compensated by vectors of the reciprocal aperiodic nonlinear
lattices \cite{Shutov1, Shutov2}.
For example, one can produce four-mode entangled states with wave lengths $\lambda_1 = \SI{2.129}{\micro\metre}$; $\lambda_2 = \SI{1.127}{\micro\metre}$; $\lambda_3 = \SI{0.709}{\micro\metre}$ and $\lambda_4 = \SI{0.532}{\micro\metre}$ if $\mathrm{LiNbO}_3$ ANPC with lattice periods from $\SI{8.0}{\micro\metre}$ to $\SI{21.2}{\micro\metre}$ is pumped by radiation with wavelength $\lambda_p = \SI{1.064}{\micro\metre}$.
It should be noted that the considered process was recently implemented in the scheme with two consistently located nonlinear photon crystals out of the resonator (cascade processes) \cite{Suchovski} where the photon pair spectrum at a frequency over the pump frequency was investigated.
 
 Quantum properties of the  process  under consideration were comprehensively studied in works \cite{Saygin, Tlyachev,Saygin2}, where
 existence of four-partite entanglement was established. It should be noted that quantum correlations exist not only between photons at low frequencies and between photons at the highest frequencies in relation to pump frequency, but also between low-frequency and high-frequency photons.
Some references to other methods of generation of multipartite entangled light fields are given in \cite{Chirkin}.

 In the case of collinear wave interaction geometry,  beams after exiting ANPC have to be spatially divided, for example, by a
dispersive prism.  We denote beam amplitudes operators as
$\hat {A} _ {j} (\vr, l) $ where the index $j$ corresponds to number of frequency, $l$ is the length of the crystal, and vector $\vr$ lies in the plane perpendicular to the propagation direction.

 In Fig.~\ref{fig:ghost-imaging} the   studied object  $O$ with
  transmission coefficient $T(\vr_{1}) $ is lit with the radiation of frequency $ \omega_{1}$. The detector $D_1$ registers
the light field of the entire beam, and, as noted above, its output does not contain information on spatial transparency distribution of the object. 
At the same time, information about
spatial distribution $T(\vr_ {1}) $ is extracted from the measurement of mutual intensity correlations between the object and reference arms \cite {Shapiro, Erkman, Gatti}.

The Fourier transform $\hat{a}_{j}(\vq)$ ($\vq$ is the transverse wave vector) of operators $\hat {A}_{j}(\vr)$ at the entrance and at the exit of the  nonlinear crystal are related by the expression \cite{Saygin}
\begin{equation}
\label{eqn:8}
\hat{\mathbf {a} } (\vq, l) = Q (\vq, l) {\hat{\mathbf {a} } } _ {0} (\vq, 0),
\end{equation}
where $Q$ is the $4$x$4$ matrix, elements $Q_{ mn}$ of which describe
conversion of the field from frequency $\omega_ { n}$ to frequency
$\omega_ { m}$. Diagonal elements of this matrix describe the
conversion of the operator of the field at frequency $\omega_ { m}$.
The expression for the matrix \(Q \) and its properties are described in \cite{Saygin}.
In~\eqref{eqn:8} operators $\hat{\mathbf{a}}$ and ${\hat{\mathbf {a}}}_{ 0}$ are columns of two-dimensional  Fourier transforms of spatial creation and annihilation  photon operators  at the exit and at the
entrance of the nonlinear crystal, respectively. The columns are of the form
$\hat{\mathbf{a}} = (\hat{a}_{1}, \hat{a}_2^{\dag}, \hat{a}_{3}, \hat{a}_{4}^{\dag})^T$,
where symbol $T$ means transposing and $\hat{a}_1 = \hat{a}_1(\vq, l)$, $\hat{a}_2^\dag = \hat{a}_2^\dag(-\vq, l)$, $\hat{a}_3 = \hat{a}_3(\vq, l)$, $\hat{a}_4^\dag = \hat{a}_4^\dag(-\vq, l)$.

The amplitude operators of  fields in the plane of detectors are related to the operators at the crystal's
exit by the following relation:
\begin{equation}
\label{eqn:11}
{\hat{B}}_{m}(\vr_{m}) = \int H_{m}(\vr_m, \vrho){\hat{A}}_{m}(\vrho, l) \mathrm{d}\vrho,
\end{equation}
where
$H_{m}(\vr_m, \vrho)$ is the medium response function related to propagation of the radiation from the crystal to the detector in the $m$th  arm.
Note that in \eqref{eqn:11} we neglected additive operator terms associated with possible losses in the imaging systems (that are necessary to satisfy the commutation relations), as they do not correlate with $A_m(\vrho)$ and do not contribute to the intensity correlations of interest (see also \cite{Gatti}).

\section{Correlations of ghost images}
\pst
In connection to the stated problem,  calculation of the following parameters is necessary:
\begin{itemize}
\item the mean value of the intensity operator 
\begin{equation}
\label{eqn:14}
\left\langle {\hat{I}}_{m}(\vr) \right\rangle = \left\langle {\hat{B}}_{m}^{\dagger}(\vr_{m}){\hat{B}}_{m}(\vr_{m})\right\rangle;
\end{equation}
\item the mutual intensity fluctuation operator
\begin{equation*}
\hat{G}_{1i}(\vr_{1}, \vr_{i}) = \hat{I}_{1}(\vr_1)\hat{I}_{i}(\vr_i)  - \left\langle {\hat{I}}_{1}( \vr_{1}) \right\rangle \left\langle {\hat{I}}_{i}(\vr_{i}) \right\rangle,
\end{equation*}
that is actually the operator of the GI,
as its mean value contains information on the object;
\item finally, the GI correlation function that is given by
\begin{equation}
\label{eqn:16}
G_{1i1j}(\vr_{1}, \vr_{i}, \vr'_1, \vr_{j}) =
\left\langle
\hat{G}_{1i}(\vr_{1}, \vr_{i})
\hat{G}_{1j}(\vr'_{1}, \vr_{j})
\right\rangle.
\end{equation}
\end{itemize}

Averages of operator expressions are calculated for vacuum state of the field at the entrance of the nonlinear crystal.

The fourth order correlation functions of intensities  \eqref{eqn:16} present the greatest calculation difficulty:
\begin{equation*}
\left\langle {\hat{I}}_{1}{\hat{I}}_{i}{\hat{I}}_{1}{\hat{I}}_{j} \right\rangle = \left\langle {\hat{B}}_{1}^{\dagger}{\hat{B}}_{1}{\hat{B}}_{i}^{\dagger}{\hat{B}}_{i}{\hat{B}}_{1}^{\dagger}{\hat{B}}_{1}{\hat{B}}_{j}^{\dagger}{\hat{B}}_{j} \right\rangle.
\end{equation*}

Here and afterwards, arguments of the operators are omitted for brevity.

The average of the product operators in the detector planes \eqref{eqn:11} expressed through operators at the exit from a crystal has the following form:
\begin{eqnarray*}
\left\langle {\hat{I}}_{1}{\hat{I}}_{i}{\hat{I}}_{1}{\hat{I}}_{j} \right\rangle =
\int {\mathrm{d}\vrho_{1}}
\int {\mathrm{d}\vrho_{1}'}
\int {\mathrm{d}\vrho_{i}}
\int {\mathrm{d}\vrho_{i}'}
\int {\mathrm{d}\vrho_{1}''}
\int {\mathrm{d}\vrho_{1}'''}
\int {\mathrm{d}\vrho_{j}}
\int {\mathrm{d}\vrho_{j}'}
\times\\\times
H_{1}^{*}(\vr_{1}, \vrho_{1})
H_{1}(\vr_{1}, \vrho_{1}')
H_{i}^{*}(\vr_{i}, \vrho_{i})
H_{i}(\vr_{i}, \vrho_{i}')
\times\\\times
H_{1}^{*}(\vr'_{1}, \vrho_{1}'')
H_{1}(\vr'_{1}, \vrho_{1}''')
H_{j}^{*}(\vr_{j}, \vrho_{j})
H_{j}(\vr_{j}, \vrho_{j}')
\times\\\times
\left\langle {\hat{A}}_{1}^{\dagger}(\vrho_1) {\hat{A}}_{1}(\vrho'_1) {\hat{A}}_{i}^{\dagger}(\vrho_i) {\hat{A}}_{i}(\vrho'_i)
{\hat{A}}_{1}^{\dagger}(\vrho''_1) {\hat{A}}_{1}(\vrho'''_1){\hat{A}}_{j}^{\dagger}(\vrho_j){\hat{A}}_{j}(\vrho'_j) \right\rangle.
\end{eqnarray*}

The fields formed by the parametric conversion obey Gaussian statistics, therefore, we can carry out the factorization of
$\left\langle {\hat{A}}_{1}^{\dagger}{\hat{A}}_{1}{\hat{A}}_{i}^{\dagger}{\hat{A}}_{i}{\hat{A}}_{1}^{\dagger}{\hat{A}}_{1}{\hat{A}}_{j}^{\dagger}{\hat{A}}_{j} \right\rangle$ using Wick's theorem \cite{Lifshits}.
According to the theorem, the mean of the product of any number of bosonic creation and annihilation operators over vacuum state is equal to the sum of products of all possible averages of operator pair products, where in each pair factors are ordered the same way as in the initial product.
Therefore, the product under discussion is transformed to the sum of terms, each of which is a
product of four average products of a pair of operators:
\begin{multline*}
\left\langle {\hat{A}}_{1}^{\dagger}{\hat{A}}_{1}{\hat{A}}_{i}^{\dagger}{\hat{A}}_{i}{\hat{A}}_{1}^{\dagger}{\hat{A}}_{1}{\hat{A}}_{j}^{\dagger}{\hat{A}}_{j} \right\rangle =
\left\langle \hat{A}_1^\dag \hat{A}_1 \right\rangle
\left\langle \hat{A}_i^\dag \hat{A}_i \right\rangle
\left\langle \hat{A}_1^\dag \hat{A}_1 \right\rangle
\left\langle \hat{A}_j^\dag \hat{A}_j \right\rangle
+ \ldots +\\
\left\langle \hat{A}_1^\dag \hat{A}_1^\dag \right\rangle
\left\langle \hat{A}_1^\dag \hat{A}_i \right\rangle
\left\langle \hat{A}_i^\dag \hat{A}_j \right\rangle
\left\langle \hat{A}_1 \hat{A}_j^\dag \right\rangle
+ \ldots +
\left\langle \hat{A}_1^\dag \hat{A}_j \right\rangle
\left\langle \hat{A}_1 \hat{A}_j^\dag \right\rangle
\left\langle \hat{A}_i^\dag \hat{A}_1 \right\rangle
\left\langle \hat{A}_i \hat{A}_1^\dag \right\rangle
.
\end{multline*}
To carry out this procedure, we have written a program implementing the following algorithm:
\begin{enumerate}
\item
 Operators  \({\hat{A}}_{1}^{\dagger}\), $\ldots$, \({\hat{A}}_{j}\) 
  were denoted  by numbers $0$, $\ldots$, $7$, respectively.
\item
 From all permutations of the set $\{0, \dots, 7\}$ those which satisfy the following were picked:
\begin{enumerate}
\item\label{itm:order-cond} Each pair of elements of the set obtained by permutation
is increasingly ordered.
\item\label{itm:order-cond2} The elements of the set obtained by permutation with odd serial numbers (that is, the first elements of pairs), are increasingly ordered.
\end{enumerate}
\end{enumerate}

Condition \ref{itm:order-cond} is necessary for obtained  permutations to
satisfy the Wick's theorem conditions, and condition \ref{itm:order-cond2}
is necessary and sufficient for repeated terms not to
appear during factorization.

The factorized product
$\left\langle {\hat{A}}_{1}^{\dagger}{\hat{A}}_{1}{\hat{A}}_{i}^{\dagger}{\hat{A}}_{i}\right.$ $\left.{\hat{A}}_{1}^{\dagger}{\hat{A}}_{1}{\hat{A}}_{j}^{\dagger}{\hat{A}}_{j} \right\rangle$
has $105$ terms of the form $\langle \hat{A}^\dag_k \hat{A}_{k'}\rangle$, $\langle \hat{A}_k \hat{A}^\dag_{k'}\rangle$, $\langle \hat{A}^\dag_k \hat{A}^\dag_{k'}\rangle$ and $\langle \hat{A}_k \hat{A}_{k'}\rangle$.  The result of calculation of $\left\langle {\hat{A}}_{1}^{\dagger}{\hat{A}}_{1}{\hat{A}}_{i}^{\dagger}{\hat{A}}_{i}\right\rangle$ $\left\langle{\hat{A}}_{1}^{\dagger}{\hat{A}}_{1}{\hat{A}}_{j}^{\dagger}{\hat{A}}_{j} \right\rangle$ has  $9$ terms.

To calculate averages $\langle \hat{A}^\dag_k \hat{A}_{k'}\rangle$, $\langle \hat{A}_k \hat{A}^\dag_{k'}\rangle$, $\langle \hat{A}^\dag_k \hat{A}^\dag_{k'}\rangle$ and $\langle \hat{A}_k \hat{A}_{k'}\rangle$,
operators $\hat{A}_k$ and $\hat{A}^\dag_k$ are expressed through operators   $\hat{a}_k$ and  $\hat{a}^\dag_k$ by inverse Fourier transformations. Dependence of the latter on   the creation  and annihilation operators at the crystal's entrance is given by \eqref{eqn:8}.
After these actions the factorized expression depends on the Bose operators at the entrance of the crystal, and  only averages of products of antinormally ordered operator pairs have nonzero value.

As a result of calculations, taking into account that  the detector in the object arm collects radiation from the entire beam aperture and provided that the band of transverse wave numbers of the parametric converter is much larger  than the corresponding band of the image,
we obtain the expression
\begin{equation}
\label{eqn:29}
G_{ij}(\vr_{i}, \vr_{j}) =
\left(\frac{k_1}{2 \pi f}\right)^4
|T(-\vr_i)|^2
|T(-\vr_j)|^2
\int\int \mathrm{d}\vr_1 \mathrm{d}\vr'_1
W_{ij}(\vr_1, \vr_i, \vr'_1, \vr_j),
\end{equation}
where
\begin{multline*}
W_{ij}(\vr_1, \vr_i, \vr'_1, \vr_j) = \int {\mathrm{d}\vR} \int {\mathrm{d}\vd} \int {\mathrm{d}\vR'} \int {\mathrm{d}\vd'}
\exp\left(-\mathrm{i} \frac{k_1}{f} (\vr_1, \vd)\right)
\exp\left( -\mathrm{i} \frac{k_1}{f} (\vr'_1, \vd') \right)
\times\\\times
\widetilde{W}_{ij}(\vR - \vd/2, \vR + \vd/2, -\vr_i, -\vr_i,
\vR' - \vd'/2, \vR' + \vd'/2, -\vr_j, -\vr_j).
\end{multline*}

The function $\widetilde{W}_{ij}$ has an extremely cumbersome form. It is determined by the sum of products of Fourier transforms of transfer coefficients $Q_{mn} (q,l)$ in the eighth degree.

Calculation of the correlation function for mutual intensity fluctuations  yields the result coinciding with   \cite{Chirkin}:
\begin{equation}
\label{eqn:32}
G_{j}(\vr_{j})
\bydef
\int G_{1j}(\vr_1, \vr_j) \mathrm{d}\vr_{1}
=
\left(\frac{k_{1}}{2 \pi f} \right)^{2}
\left| \int {Q_{(11j)}\left(\frac{k_{1}}{f}\vr_{1}\right)}\mathrm{d}\vr_{1} \right|^{2}
\left| T( - \vr_{j}) \right|^{2},
\end{equation}
where
$Q_{(11j)}(\vq) = Q_{11}(\vq) Q_{j1}^{*}(\vq) + Q_{13}(\vq)Q_{j3}^{*}(\vq)$.

Expressions \eqref{eqn:29}, \eqref{eqn:32} have been obtained for the  case when in the arms of the set-up in Fig.~\ref{fig:ghost-imaging} 
lenses  are used (see also \cite{Gatti, Chirkin}). The object and detector $D_1$  are located
in the focal regions of a lens. In the reference arms lenses are located at double focal length from detectors $D_j$ and the crystal. Lenses are not portrayed in Fig.~\ref{fig:ghost-imaging}.

\section{Interpretation of acquired ghost images}
\pst
We use a two-dimensional array of detectors as a measuring device.
The output of each detector is proportional to incoming luminous flux. The values obtained at each correlator's output, denoted as $\xi(\vr)$, can be represented by the effect on the measuring transducer (MT) on the input signal $f(\vr) \sim |T(-\vr)|^2$. In this article we consider piece-wise constant images, i.\,e. transparency is constant within each pixel. The algorithm of image interpretation ought to give a maximally accurate estimate of the original image $f(\vr)$ from the acquired data $\xi(\vr)$.

Let us represent the measurement model by
$\xi(\vr) = (A f)(\vr) + \nu(\vr)$,
where $f(\vr)$ is an a~priori unknown vector describing the transparency distribution of the measured object;
$A$ is the matrix describing the formation and acquisition of ghost images: the matrix element $A_{ij}$ is equal to the mean signal of $i$-th detector for unit transparency of the $j$-th object element and zero transparency of other ones;
$\nu(\vr)$ is the noise with zero mean value, corresponding to lack of systematic measurement error, and with covariance matrix
$(\Sigma_{\nu})_{ij} = \langle \nu(\vr_{i}) \nu(\vr_{j}) \rangle$
that models distortions obtained during flux measurements using the MT.
The vector $\xi(\vr)$ represents the results of flux measurements.
The dimension of vector $f(\vr)$ is given by the number of pixels in the image, and the dimension of the vector $\xi(\vr)$ is given by the number of the pixels in detector arrays.

Operators $A$ and $\Sigma_{\nu}$ are related to correlation functions. Since the measurement set-up uses correlators that measure correlations between the first arm and the other arms, MT's effect on the image is given by a block matrix consisting of three blocks representing the correlators' outputs~--- the correlations of the object arm and the reference arms:
\begin{equation}
\label{eqn:33-34}
A =
\begin{pmatrix}
B_2 C_2\\
B_3 C_3\\
B_4 C_4
\end{pmatrix},
\end{equation}
where under conditions given in derivation of \eqref{eqn:29}, \eqref{eqn:32}
matrices $C_j$ are proportional to the identity matrices multiplied by (up to a factor defined by the choice of the measurement units) the pixel size and the factor in front of
$|T(\vr_{i})|^{2}$ in the expression for $G_{j}$ \eqref{eqn:32},
and matrices $B_j$ describe the detectors:
the matrix element $(B_j)_{pk}$ is equal to the response of the detector at the $p$-th position in the $j$-th reference arm to the unit luminance of the $k$-th pixel and a zero luminance of the other ones.

The noise covariance matrix is a block matrix as well:
\begin{equation}
\label{eqn:35}
\Sigma_\nu =
\begin{pmatrix}
B_2\Sigma_{22}(f)B_2^* & B_2\Sigma_{23}(f)B_3^* & B_2\Sigma_{24}(f)B_4^*\\
B_3\Sigma_{32}(f)B_2^* & B_3\Sigma_{33}(f)B_3^* & B_3\Sigma_{34}(f)B_4^*\\
B_4\Sigma_{42}(f)B_2^* & B_4\Sigma_{43}(f)B_3^* & B_4\Sigma_{44}(f)B_4^*
\end{pmatrix}.
\end{equation}
The element of the block $\Sigma_{ij}$ with indices $k$, $k'$ is equal (up to a factor defined by the choice of the measurement units) to the value of integral of $G_{ij}$ over values of $\vr_{i}$ that belong to the $k$-th pixel and over values of $\vr_{j}$ that belong to $k'$-th pixel.

The objective of the measurement is to obtain the most accurate estimate of $U f$, where operator $U$ describes the ideal measuring transducer, using the measurement data obtained as described above.
As show in \cite{Pytyev,Balakin}, the linear estimate with the least mean squared error (MSE) is
\begin{equation}
\label{eqn:38}
R_* \xi = U (A^* \Sigma_{\nu}^- A)^- A^* \Sigma_{\nu}^- \xi,
\end{equation}
where ${}^-$ denotes matrix pseudoinverse and $R_*$ is the reduction operator defined by \eqref{eqn:38}.
Note that the \emph{mathematical} reduction method is applicable for any other light source~--- the peculiarity of the described light source is the form of covariance operator $\Sigma_{\nu}$ that affects step~\ref{itm:cov-op-estimation} of the processing algorithm given below.

Synthesis of such an estimate is possible if the condition $U (I - A^- A) = 0$ is satisfied, where, as noted above, $A$ described the \emph{real} measuring device, and $U$ describes the \emph{ideal} measurement device with the impulse response function required by the researcher and, consequently, \emph{any required resolution}, if this condition is satisfied.
However, usually the higher the desired resolution of the ideal measuring device compared to the real one, the higher the MSE of the synthesized estimate.
The specific dependency of the resolution, defined as the maximal rank of $U$ for which estimation of $U f$ with MSE below given is possible (effective rank, see \cite[ch.~8]{Pytyev}), depends first and foremost on $B_j$ and the covariance operator of the noise unrelated to ghost imaging.
For example, in the case of ideal detectors and white noise this dependency is a linear one --- quadruple relaxation of MSE requirement allows to reduce the pixel size by two times.

\section{Computer modelling results}
\pst
For simplicity we consider the case of ideal identical detectors in all the reference arms, $B_2 = B_3 = B_4 = U = I$.

The covariance matrix \eqref{eqn:35} depends on the unknown input signal.
Therefore, the measurement reduction algorithm is an iterative one, with each iteration consisting of three steps:
\begin{enumerate}
\item estimation of image using \eqref{eqn:38} for $\Sigma_{\nu}$ corresponding to the image with constant transparency and $A$ given by \eqref{eqn:33-34} (on the first iteration) or for
$\widetilde{\Sigma}_{\nu} =
\begin{pmatrix}
\Sigma_{\nu, \textnormal{e}} & 0\\
0 & \kappa I
\end{pmatrix}$,
$\widetilde{A} =
\begin{pmatrix}
A\\
I
\end{pmatrix}$,
$\widetilde{\xi} =
\begin{pmatrix}
\xi\\
\hat{f}_\textnormal{e}
\end{pmatrix}$,
where $A$ is given by \eqref{eqn:33-34}, $\hat{f}_{\textnormal{e}}$ is the image estimate obtained during previous iteration, $\kappa$ is an arbitrary positive number and $\Sigma_{\nu, \textnormal{e}}$ is the estimate of the covariation matrix obtained in the last step of the previous iteration (on later iterations);
\item orthogonal projection of the result onto $[0, 1]^{\dim f}$ to take into account that transparency takes values in $[0, 1]$;
\item \label{itm:cov-op-estimation} substitution of $f$ in \eqref{eqn:35} by the obtained estimate to refine the estimate of $\Sigma_{\nu}$, taking into account that $\hat{f}_e$ and $\xi$ are \emph{correlated}.
\end{enumerate}

\begin{figure}[tbh!]
\centering
\subfigure[\label{fig:sample-src}]{\includegraphics[scale=1]{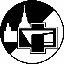}}
\subfigure[\label{fig:sample-arm2}]{\includegraphics[scale=1]{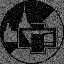}}
\subfigure[\label{fig:sample-arm3}]{\includegraphics[scale=1]{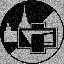}}
\subfigure[\label{fig:sample-arm4}]{\includegraphics[scale=1]{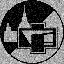}}
\subfigure[\label{fig:sample-sum}]{\includegraphics[scale=1]{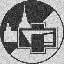}}
\subfigure[\label{fig:sample-estimate}]{\includegraphics[scale=1]{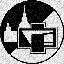}}
\caption{Images of the object acquired using $64$x$64$ pixels array and the result of their reduction: \subref{fig:sample-src}~the original image; \subref{fig:sample-arm2}--\subref{fig:sample-arm4} acquired ghost images, \subref{fig:sample-sum}~superposition of ghost images~\subref{fig:sample-arm2}--\subref{fig:sample-arm4}, \subref{fig:sample-estimate}~the result of reduction of ghost images, taking their correlations into account}
\label{fig:sample}
\end{figure}

The computer modelling results presented in Fig.~\ref{fig:sample} illustrate an application of above-described processing of MGI.
One can see that the image shown in Fig.~\ref{fig:sample-estimate} has better quality than the ghost ones (figures~\ref{fig:sample-arm2}--\ref{fig:sample-arm4}) and the one (Fig.~\ref{fig:sample-sum}) obtained by summing them up.
The simulation was carried out for the following values of parameters:
beam wave numbers $k_1 = 6~\cdot~10^{4}~\si{\per\centi\metre}$, $k_3 = 1.7~\cdot~10^5~\si{\per\centi\metre}$,
ANPC parameter $\beta = 10~\si{\per\centi\metre}$,
ANPC parameter $\xi = \gamma / \beta = 0.4$,
dimensionless ANPC thickness $\zeta = \beta l = 6$.
The signal-to-noise ratio for the entire image is $4.6$ times higher than the best signal-to-noise ratio for the ghost images themselves, and $7.7$ times higher than for their sum if only fluctuations related to ghost imaging are present.
The signal-to-noise ratio of the sum of images is different from that of each image due to fluctuations of the images.
Therefore, fluctuations in reference arms are partially suppressed by summation, although not to the same degree as if the images were completely uncorrelated. The theoretical change of signal-to-noise ratio in this case is
\begin{equation*}
(c_2 + c_3 + c_4)^2
\left(
\begin{pmatrix}
1 & 1 & 1
\end{pmatrix}
C
\begin{pmatrix}
1\\
1\\
1
\end{pmatrix}
\right)^{-1}
C_{j_*, j_*}
c_{j_*}^{-2}
\approx 0.6,
\end{equation*}
where, in accordance with formulas for mean and variance of correlated random variables,
the first factor is the squared product of unit coefficients with which the images are added and factors $c_2 = {2.73}$, $c_3 = {4.98}$, $c_4 = {5.11}$ in front of identity matrices in $C_j$ in \eqref{eqn:33-34},
the second factor is the product of the row vector of unit coefficients, the matrix
$C =
\begin{pmatrix}
{0.19} & {0.13} & {0.12}\\
{0.13} & {0.11} & {0.11}\\
{0.12} & {0.11} & {0.11}
\end{pmatrix}$
composed of image covariances, and and the column vector of the coefficients,
$j_*$ is the number of the ghost image with the best S/N ratio (in this case, the last one).
It is in good agreement with the value $4.6/7.7 \approx 0.6$ obtained by computer modelling.
In this case summation does not increase the S/N ratio, because large S/N ratio of the first GI offsets the partial suppression of fluctuations by summation.

\section{Conclusion}
\pst
To conclude, we emphasize that the a priori information used for image interpretation consists of  ghost image correlation functions. In the considered scheme these correlations are due to the entangled light states  produced by the multipartite nonlinear optic process. The computer modelling that was carried out in accordance with developed algorithm showed high efficiency of the proposed method both in sense of improving the image quality and in sense of noise suppression: poorly distinguishable images become easily recognizable after such processing.

An advantage of the suggested MGI scheme over a standard one (without multiplexing) is that several images are obtained simultaneously and frequencies of the object beam and the restoring beams can differ by several octaves.
The considered coupled parametric interactions give opportunity to obtain entangled field states between the telecommunication wave length  (about \SI{1.5}{\micro\metre}) and the wavelength (about \SI{0.8}{\micro\metre}) which is of interest to recording of optical information. Quantum fields with the specified properties can not be obtained by means of one conventional parametric process and beam splitters.

\section*{Acknowledgments}
\pst
The authors are grateful for useful discussion of the work to M.\,Yu.\,Say\-gin and A.\,P.\,Shku\-ri\-nov.
 
This work was supported by RFBR grant N 14-02-00458.


\end{document}